\documentclass[11pt,a4paper,twoside]{article}
\usepackage{mashs08}
\usepackage{graphicx}
\usepackage{amssymb}
\usepackage[latin1]{inputenc}
\usepackage{url}

\begin{document}

\thispagestyle{empty}

\title{Mining a medieval social network by kernel SOM and related methods}

\author{{\bf Nathalie Villa$^{1,2}$, Fabrice Rossi$^3$ \& Quoc Dinh Truong$^4$}\\ $^1$IUT de Carcassonne, Université de Perpignan, Perpignan. France \\ $^2$Institut de Mathématiques de Toulouse, Université de Toulouse, Toulouse, France\\{\bf nathalie.villa@math.univ-toulouse.fr}\\$^3$INRIA, Projet AxIS, Rocquencourt, France\\{\bf fabrice.rossi@inria.fr}\\$^4$IRIT, Université de Toulouse, Toulouse, France\\{\bf truong@univ-tlse2.fr}}
\abstract{This paper briefly presents several ways to understand the organization of a large social network (several hundreds of persons). We compare approaches coming from data mining for clustering the vertices of a graph (spectral clustering, self-organizing algorithms\ldots) and provide methods for representing the graph from these analysis. All these methods are illustrated on a medieval social network and the way they can help to understand its organization is underlined.}
\keywords{social network, large graphs, SOM algorithm, graph drawing, clustering, spectral clustering, heat kernel}

\maketitle

\section{Introduction}

A large number of practical applications can be modeled through what is commonly called a ``complex network''. Complex networks are relational data, that appear in World Wide Web studies, in social networks or in biological studies (genes, proteins, metabolites interaction networks) for example.

This work is based on a historical database built from the archives of Lot, a small region in South West of France. This database has already been presented in \cite{boulet_etal_MASHS2007}: in a tiny geographical location around Castelnau-Montratier, a large documentation has been collected (see \cite{hautefeuille_T1998} for a complete presentation). This documentation, made of about 1000 agrarian contracts (available at \url{http://graphcomp.univ-tlse2.fr}) is a very precious source of information about the peasants' usual life in the middle ages where most of the written documents were concerned by the well-educated part of the population. All the contracts are agrarian transactions: they mention the name of the involved peasant (or the peasants), the names of the lord and the notary to whom the peasants are related, some of the neighbors of the peasants and various other informations (such as the type of transaction, the location, the date, and so on). All the studied transactions were written between 1260 and 1340 that is, just before the Hundred Years' War but others concerned the period just after this War.

From this database, a relational network is built following the advices provided by the historians (see \cite{boulet_etal_N2008}). This social network is described by a \emph{weighted graph} with 615 vertices (the peasants) and 4193 edges standing for the relations between them. The edges are weighted by the number of relations found between two given peasants. The obtained graph is described in \cite{boulet_etal_N2008} where it is analyzed through the comparison of an algebraic study and of a SOM algorithm. The collaboration between mathematicians, computer scientists and historians intends to provide several tools for historians to understand this complex and large network. A part of the methods developed are coming from statistics and data mining and will be reviewed and illustrated in this paper. Complementary material could be found in \cite{boulet_etal_N2008} and complementary studies of the database are available in \cite{boulet_jouve_RNTI2007,boulet_etal_MASHS2007}.

The paper is organized as follows: Section~\ref{clustering} presents the problem of clustering the vertices of a large graph and explains how this problem can help to understand the structure of the graph. Several methods are reviewed and some of them, coming from what is called \emph{spectral clustering}, are illustrated on the medieval database. Section~\ref{drawing} explains how this clustering can be used to provide a simplified representation of the graph. This leads us to use organization algorithms designed for graphs in order to classify and organize simultaneously the vertices of the graph: kernel SOM, described in section~\ref{ksom}, targets such a dual objective. Finally, Section~\ref{drawing_som} intends to represent the whole graph from this final organizing map. Examples of insights on the data obtained by the reviewed methods are given at each step of the analysis.

\section{Clustering the vertices of the graph\label{clustering}}

Large graphs representing complex networks are not easy to understand. One way to simplify them, in order to underline the main tendances of their structure, is to find dense subgraphs that have few connections to each others. As emphasized by \cite{newman_girvan_PRE2004},
\begin{quotation}
	{\it reducing [the] level of complexity [of a network] to one that can be interpreted readily by the human eye, will be invaluable in helping us to understand the large-scale structure of these new network data.}
\end{quotation}

But such a clustering of the vertices of a graph is facing the problem of relational data: there is no {\it a priori} distance between two vertices and thus classical clustering algorithms, such as, e.g., $k$-means, cannot be directly used. A recent survey on clustering methods adapted to graphs is provided in \cite{schaeffer_CSR2007}. Clustering the vertices of a graph is commonly addressed by the use of a dissimilarity between vertices or by mapping the graph on an euclidean space; then usual data mining tools can be used to find a convenient clustering. Recently, {\it spectral clustering} became a successful method among this kind of methodologies (see \cite{vonluxburg_SC2007} for a very exhaustive tutorial on this subject): spectral clustering uses the properties of the {\it Laplacian} of the graph to understand its structure. Given a weighted graph ${\cal G}$ with vertices $V=\{x_1\ldots,x_n\}$ and edges weighted by $(w_{i,j})_{i,j=1\ldots,n}$ ($w_{i,j}=w_{j,i}$ and $w_{i,i}=0$), the Laplacian is the matrix $L$ such that $L_{i,j}=\left\{\begin{array}{ll}
	-w_{i,j} & \textrm{ if } i\neq j,\\
	d_i=\sum_{j=1}^n w_{i,j} & \textrm{ if } i=j.
\end{array}\right.$
The Laplacian appears as a very convenient tool for understanding the graph as its eigenvalue decomposition is directly related to the min cut problem (``How to find a partition of the vertices that minimizes the number of cuts in the graph ?'', see \cite{vonluxburg_SC2007}) and to the problem of finding perfect communities, i.e., complete subgraphs those vertices have exactly the same neighbors (see \cite{boulet_etal_N2008,vandenheuvel_pejic_AOR2001}). More precisely, spectral clustering uses the eigenvectors associated with the smallest eigenvalues of the Laplacian to map the graph on an Euclidean space where a $k$-means algorithm is performed.

But \cite{boulet_etal_N2008} notes that the spectral clustering method gives equal weights to the first $p$ eigenvectors of the Laplacian, whereas the smaller the eigenvalue is, the more important the corresponding eigenvector is. Moreover, only the first $p$ eigenvalues are used and, hence, this approach does not use the entire structure of the graph. To avoid these problems, one can use a regularized version of the Laplacian: the heat kernel (also called the diffusion kernel). The diffusion matrix of the graph ${\cal G}$ for the parameter $\beta>0$ is $D^\beta=e^{-\beta L}$ and the diffusion kernel is the function $K_\beta : (x_i,x_j)\in V\times V \rightarrow D^\beta_{i,j}$. This diffusion kernel has been intensively studied and used through the past years (see \cite{chung_SGT1997,kondor_lafferty_ICML2002,smola_kondor_COLT2003,scholkopf_tsuda_vert_KMCB2004,vert_kanehisa_B2003}, among others). One of its main desirable properties comes from Aronszajn's Theorem \cite{aronszajn_TAMS1950} that states that there is a reproducing kernel Hilbert space (RKHS), ${\cal H}_\beta$, called the feature space, and a mapping function, $\phi_\beta : V \rightarrow {\cal H}_\beta$ such that for all $i,j$, 
\begin{equation}
	\label{kernel_trick}
	\langle \phi_\beta(x_i),\phi_\beta(x_j)\rangle_{{\cal H}_\beta}=K_\beta(x_i,x_j).
\end{equation}
This last equation is commonly known as {\it kernel trick} and means that $K_\beta$ is simply a scalar product between images by $\phi_\beta$ of the vertices of the graph. Similarly as spectral clustering, a $k$-means algorithm can be performed on this mapping. This method is known under the name {\it kernel $k$-means} (see \cite{shawetaylor_cristianini_KMPA2004,filippone_etal_PR2008,dhillon_etal_ICKDD2004}).

On a practical point of view, partitions coming from spectral clustering and kernel $k$-means cannot be directly compared by the way of $k$-means error because the mapping of the graph is not the same: the underlined metrics are not comparable. The same occurs for partitions coming from kernels with different values of $\beta$. To enable a comparison between all these partitions, we use a quality measure introduced by \cite{newman_PRE2003}, the \emph{$q$-modularity}, $Q_\textrm{modul}=\sum_{j=1}^k (e_j-a_j^2)$ where $k$ is the number of clusters, $e_j$ is the fraction of edges in the graph that connect two vertices in cluster $j$ and $a_j$ is the fraction of the edges in the graph that connect to one vertex in cluster $j$. This criterion does not depend on a mapping or a dissimilarity on the graph and is easily interpretable in terms of probability of having in/between-clusters edges: a high $q$-modularity means that vertices are clustered into dense subgraphs having few edges between them.

Table~\ref{res_clustering} summarizes the main characteristics of the partitions into 50 clusters of the medieval graph obtained by these two approaches. Obviously, both partitions share common properties: the $q$-modularity is similar and the vertices are concentrated in a few number of large clusters. More than three fourth of the vertices belong to a cluster having less than 7 vertices and the largest cluster contains more than one third of the vertices of the whole graph. In conclusion, the partitions provided by these two simple tools have to be improved.

\begin{table}[ht]
	\begin{center}
		\fbox{\small \begin{tabular}{l|c|c}
			& Spectral clustering & Kernel $k$-means\\
			&& ($\beta=0.05$)\\
			\hline
			$q$-modularity & 0.4195 & 0.4246 \\
			\hline
			Number of clusters of size 1 & 16 & 17\\
			Maximum size of the clusters & 268 & 242\\
			Median of the clusters' size & 2 & 2\\
			3{\it rd} quartile of the clusters' size & 7 & 7
		\end{tabular}}
		\caption{Details about the partitions obtained by spectral clustering and kernel $k$-means\label{res_clustering}}
	\end{center}
\end{table}

\section{Drawing the graph\label{drawing}}

\begin{figure}[ht]
	\begin{center}
		\begin{tabular}{cc}
			\includegraphics[width=5 cm]{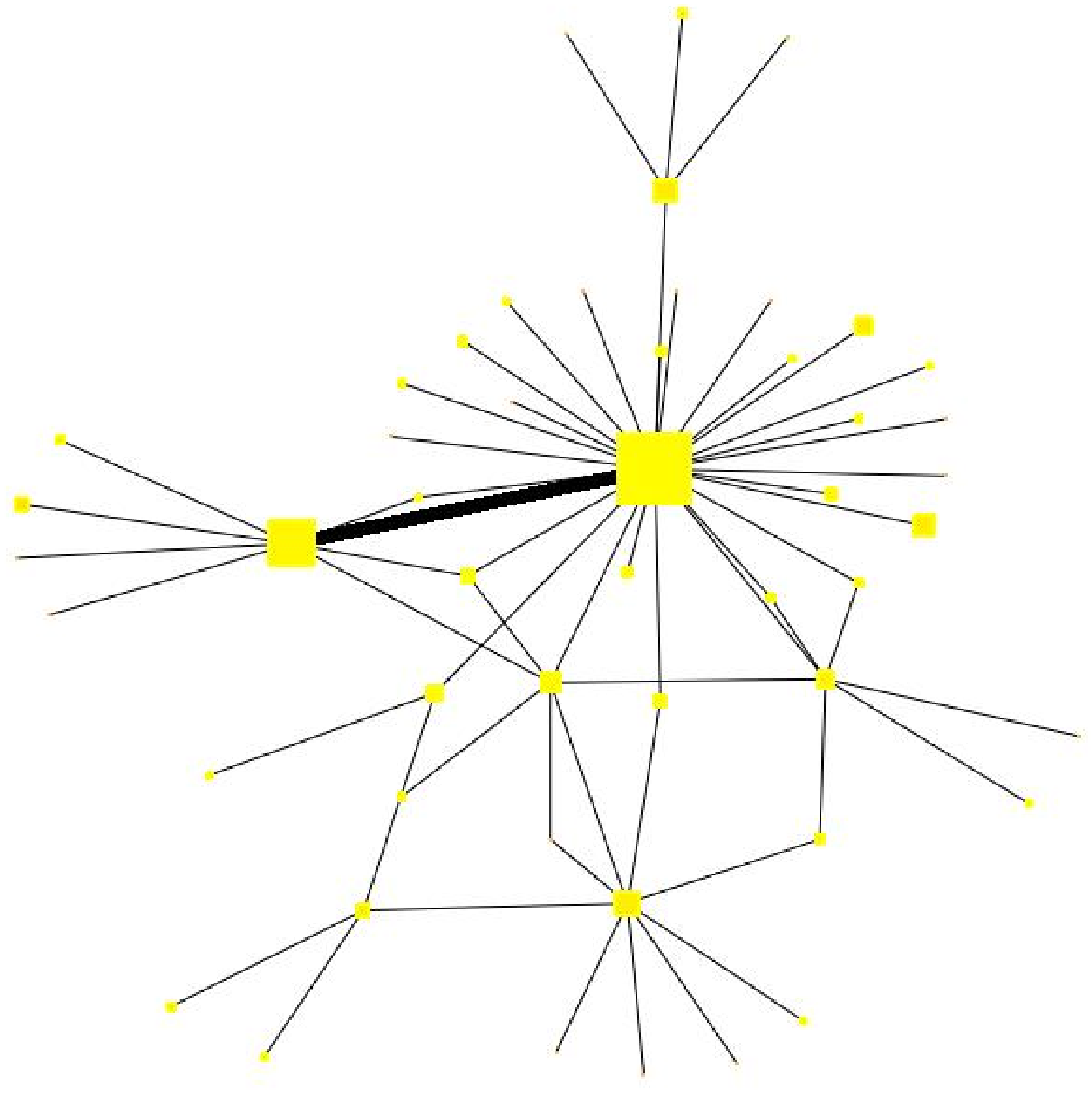} & \includegraphics[width=5 cm]{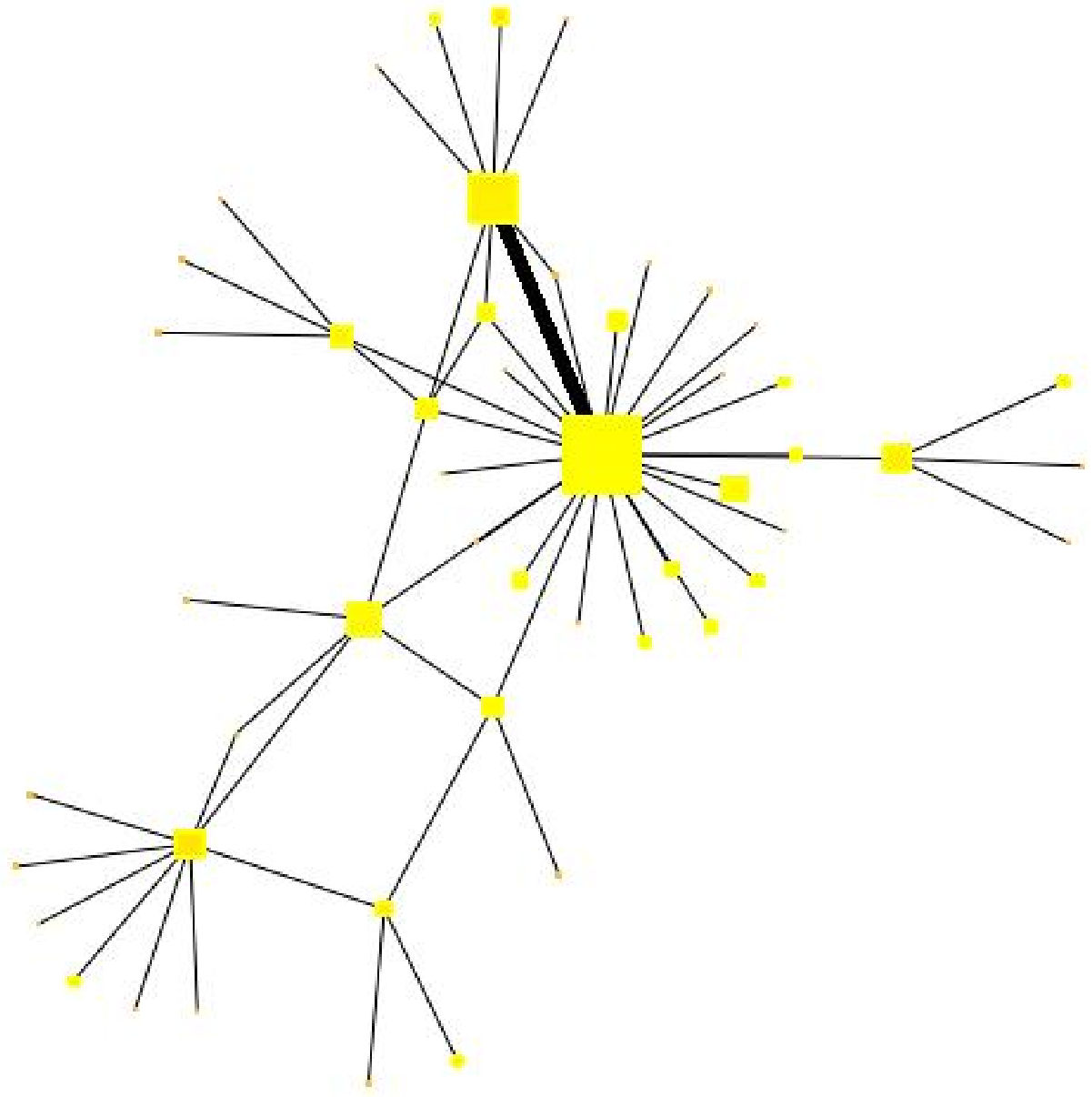}
		\end{tabular}
		\caption{Force-directed algorithms used for a simplified representation coming from spectral clustering (left) and kernel $k$-means (right)\label{figure_spectralclustering}}
	\end{center}
\end{figure}

From any partition obtained by clustering the graph, a simplified representation can be obtained by assigning a given glyph to each cluster where the surface of the glyph is proportional to the number of vertices of the given cluster. At the same time, glyphs are connected to each others by edges whose width is also proportional to the total number of weights between the vertices belonging to the two corresponding clusters. Glyphs can be spatially positioned by a \emph{force directed algorithm} that aims at providing an aesthetic representation of a graph by assigning forces amongst the edges and nodes (see \cite{fruchterman_reingold_SPE1991}). Examples of such a representation are given in Figure~\ref{figure_spectralclustering} for the two partitions described in section~\ref{clustering}\footnote{All the graph figures have been made with the free software Tulip, available at \url{http://www.tulip-software.org/}}.

Both representations share common properties that can be seen as main structural properties of the graph: the network has a star shaped structure with two main groups of central people that can be seen as a kind of ``rich club'' (see \cite{boulet_etal_N2008}). Some tiny groups are totally isolated from this two main central clusters and linked to other secondary clusters. The two major clusters are strongly linked to each others. These pictures seem to give understandable representations of the structure of the graph but, unfortunately, the two main clusters respectively contain about 250 and 100 vertices, that is, more than half of the vertices of the graph: then, these two clusters are almost as complex as the initial graph.

\section{A clustering and organizing algorithm\label{ksom}}

Several kernelized versions of SOM algorithm, that can perform simultaneously the objectives described in sections \ref{clustering} (clustering) and \ref{drawing} (representation), has been described in \cite{lau_etal_N2006}. The present paper uses a batch version of the kernel SOM (that generally converges much faster) proposed in \cite{boulet_etal_N2008,villa_rossi_WSOM2007}.

The aim of self-organizing algorithms is to project the initial data on a prior structure that is generally a grid consisting in M neurons. A neighborhood relationship is defined on the grid and the projection intends to preserve the initial topology of the data on this grid. The batch kernel SOM is simply a batch SOM performed on data that have been mapped on a RKHS; the algorithm is rewritten by the way of the kernel trick (Equation~(\ref{kernel_trick})).

This algorithm has been applied to the medieval graph with a rectangular grid of size 7$\times$7. The main characteristics of the obtained partition is summarized in Table~\ref{res_ksom}. It is compared to the partition obtained by using the batch SOM on the rows of the $k$ eigenvectors associated to the smallest eigenvalues (this last approach has been named ``spectral SOM''). 
\begin{table}[ht]
	\begin{center}
		\fbox{\small \begin{tabular}{l|c|c}
			& Spectral SOM & Kernel SOM\\
			&& ($\beta=0.05$)\\
			\hline
			$q$-modularity & 0.433 & 0.551 \\
			\hline
			Final number of clusters & 29 & 35\\
			\hline
			Number of clusters of size 1 & 11 & 13\\
			Maximum size of the clusters & 325 & 255\\
			Median of the clusters' size & 2 & 3\\
			3{\it rd} quartile of the clusters' size & 10 & 10
		\end{tabular}}
		\caption{Details about the clustering obtained by spectral SOM and batch kernel SOM\label{res_ksom}}
	\end{center}
\end{table}

\begin{figure}[ht]
	\begin{center}
		\begin{tabular}{cc}
			\includegraphics[width=3.5 cm]{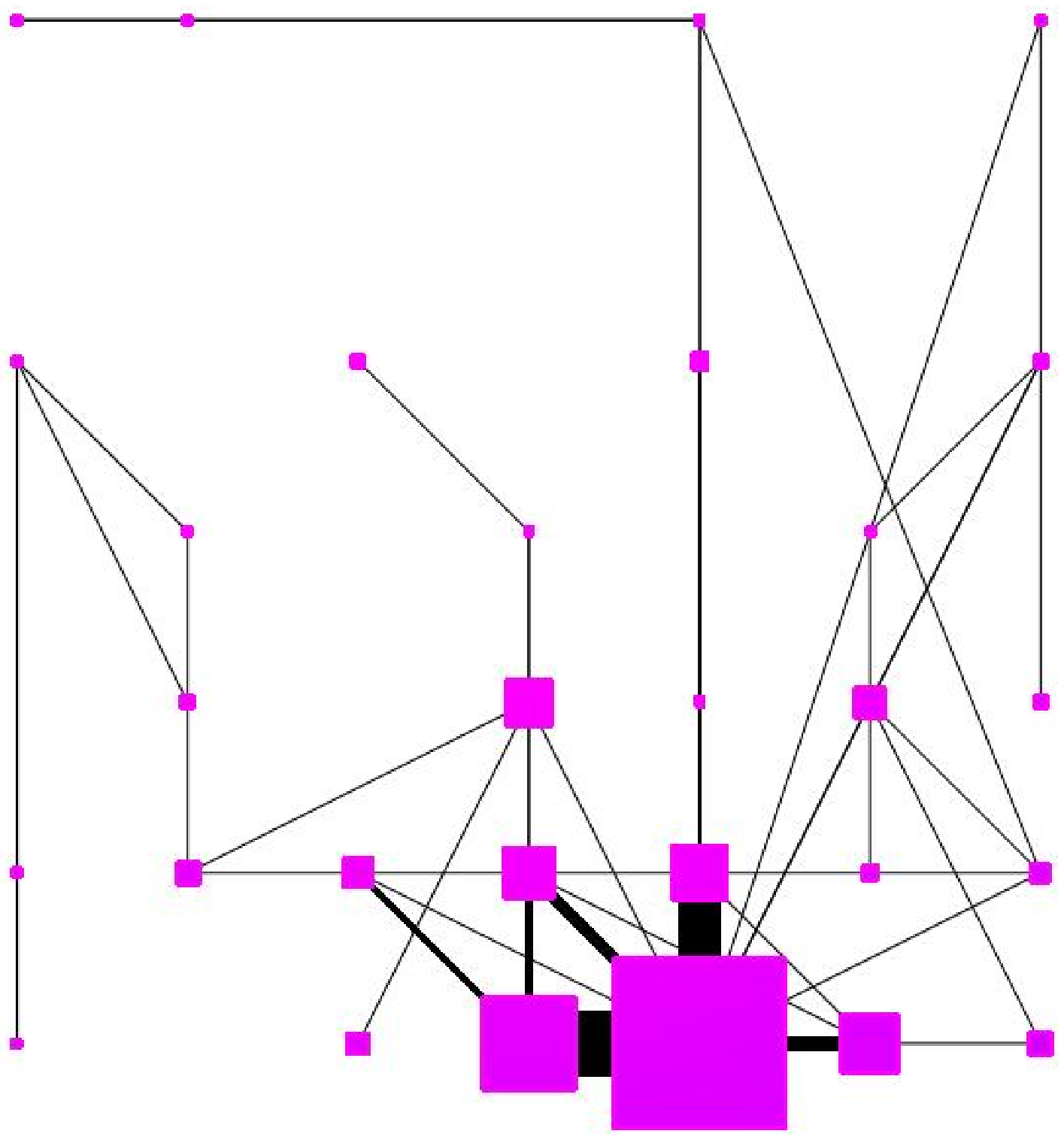}&\includegraphics[width=4.5 cm]{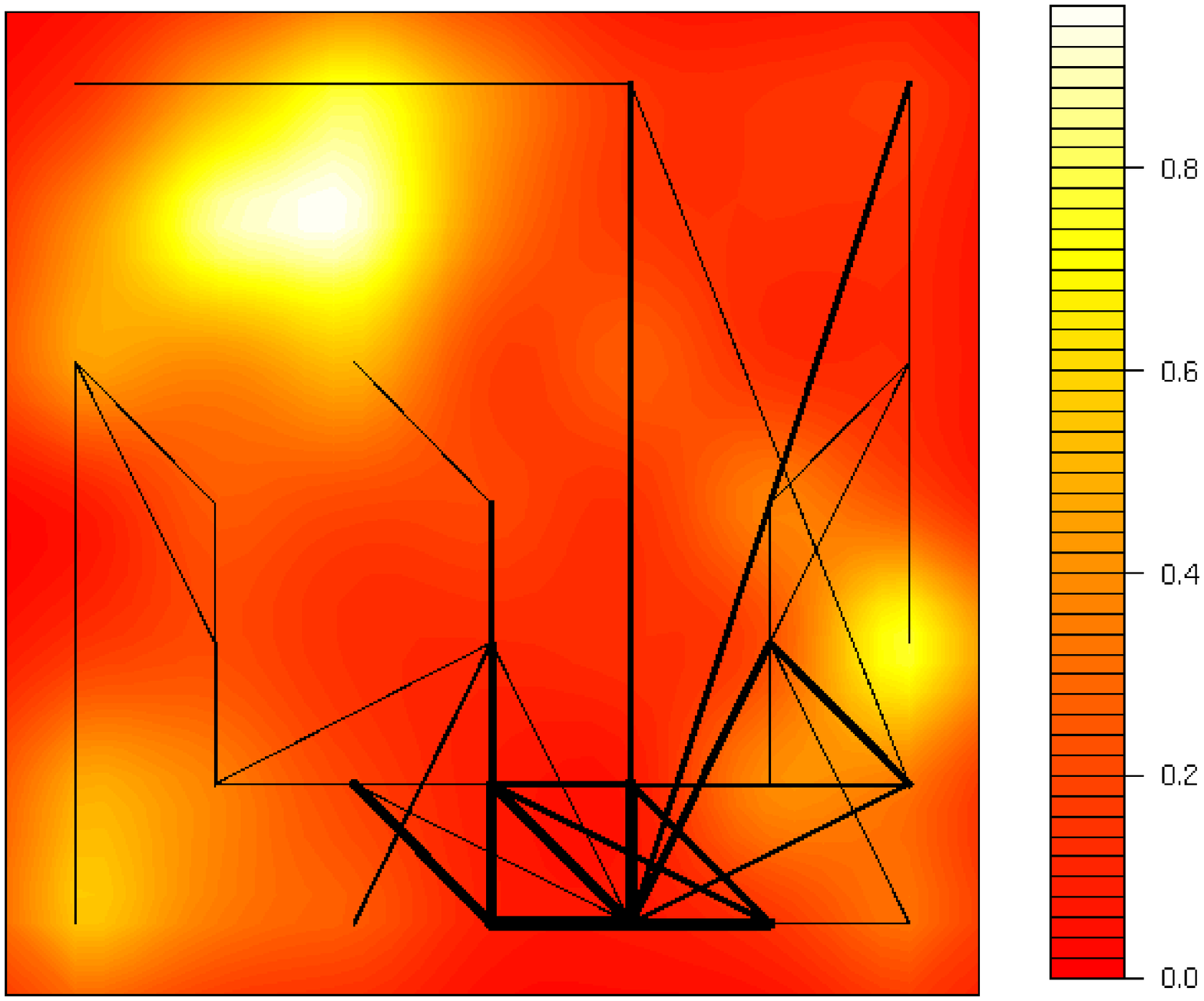}\\
			\includegraphics[width=3.5 cm]{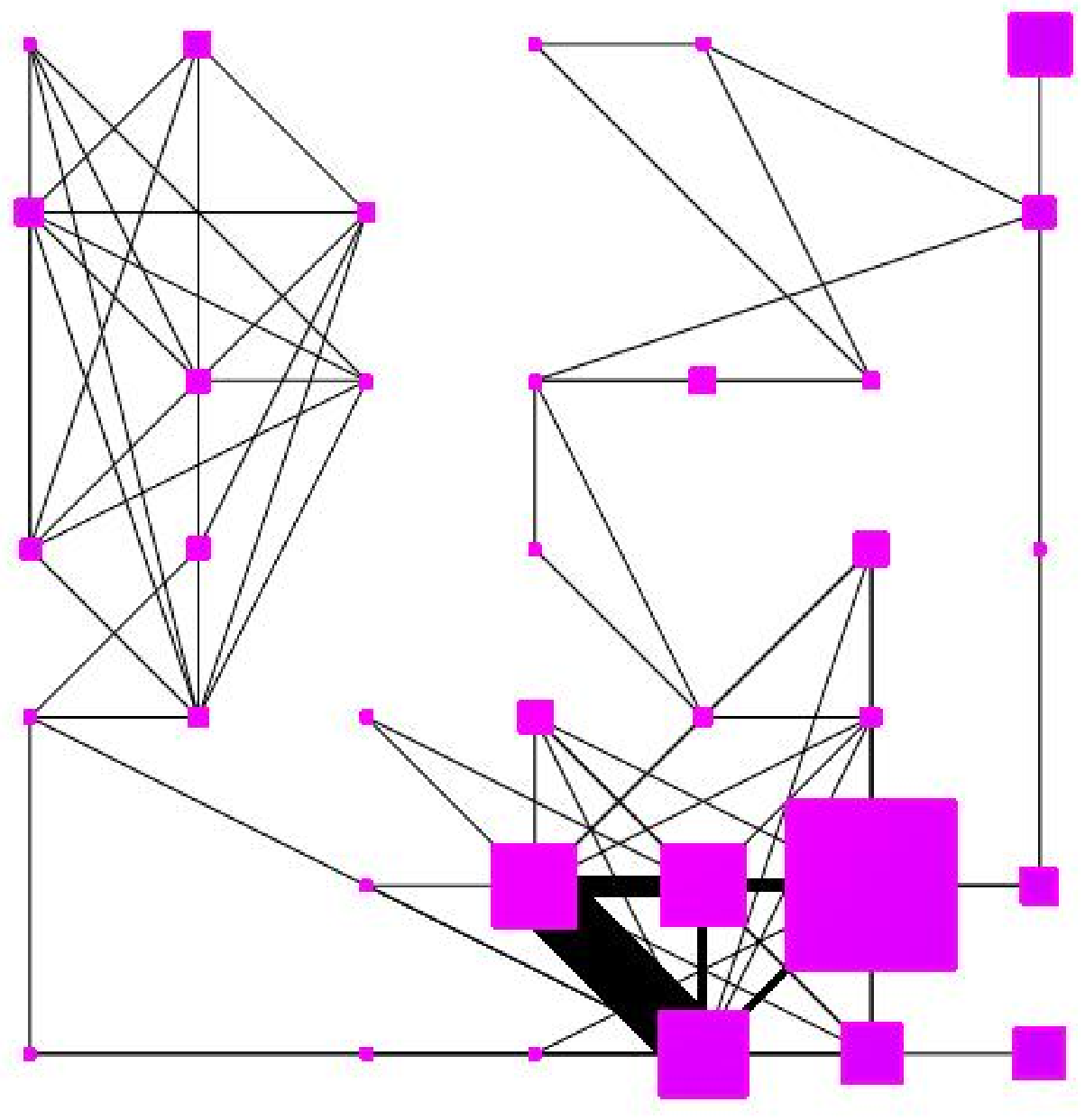}&\includegraphics[width=4.5 cm]{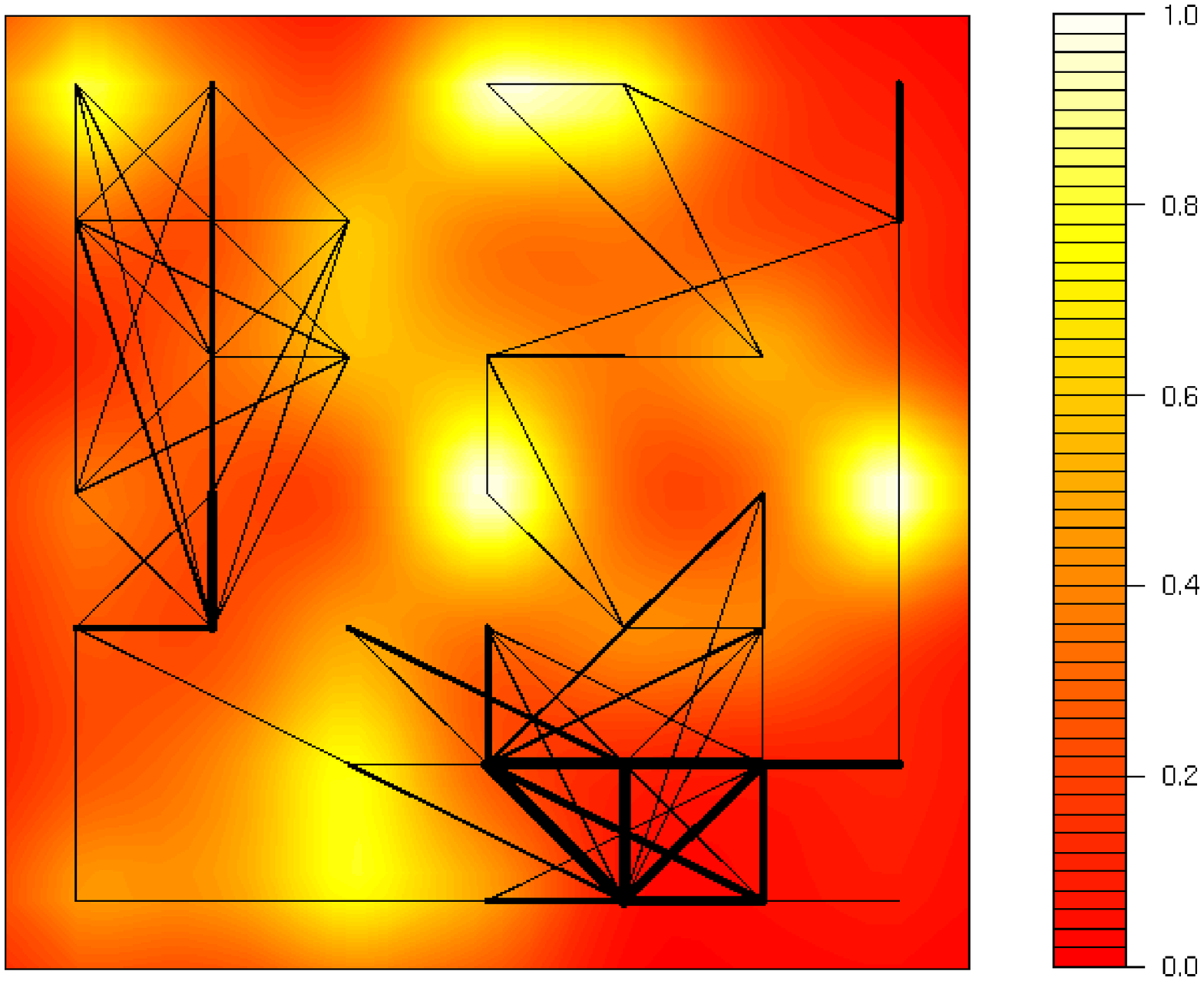}
		\end{tabular}
	\caption{Final map obtained by spectral SOM (top left) and corresponding smoothed u-matrix (top right) and final map obtained by batch kernel SOM (bottom left) and corresponding smoothed u-matrix (bottom right)\label{figure_ksom}}
	\end{center}
\end{figure}

The corresponding simplified representations, respecting the topology of the map, are provided in Figure~\ref{figure_ksom}\footnote{Colored and high quality images can be found at \url{http://nathalie.vialaneix.free.fr/maths/article-normal.php3?id_article=8}} where an additional information is given by the U-matrices (see \cite{ultsch_siemon_INNC1990}) that smoothly represent the mean distances (respectively in the $\mathbb{R}^k$ space generated by $k$ eigenvectors associated to the smallest eigenvalues and in the feature space) between the prototypes of each cluster. Clearly, kernel SOM provides better clustering than spectral SOM (larger $q$-modularity, much less vertices in the largest cluster). It also seems to be a little bit better than the spectral clustering and the kernel $k$-means (larger $q$-modularity, smaller number of tiny clusters - with less than 5 vertices).

In both cases, the simplified representations are well organized and easy to understand. Compared to Figure~\ref{figure_spectralclustering}, the representation provided by kernel SOM is very close to the one provided by a simple clustering followed by a force directed representation of the clusters: a large cluster has a central position and is surrounded by smaller clusters. But looking at the u-matrix, the map is clearly divided into three main part (top left, top right and bottom right) which, according to color levels, are distant to each others.

\begin{figure}[ht]
	\begin{center}
		\begin{tabular}{cc}
			\includegraphics[height=5 cm]{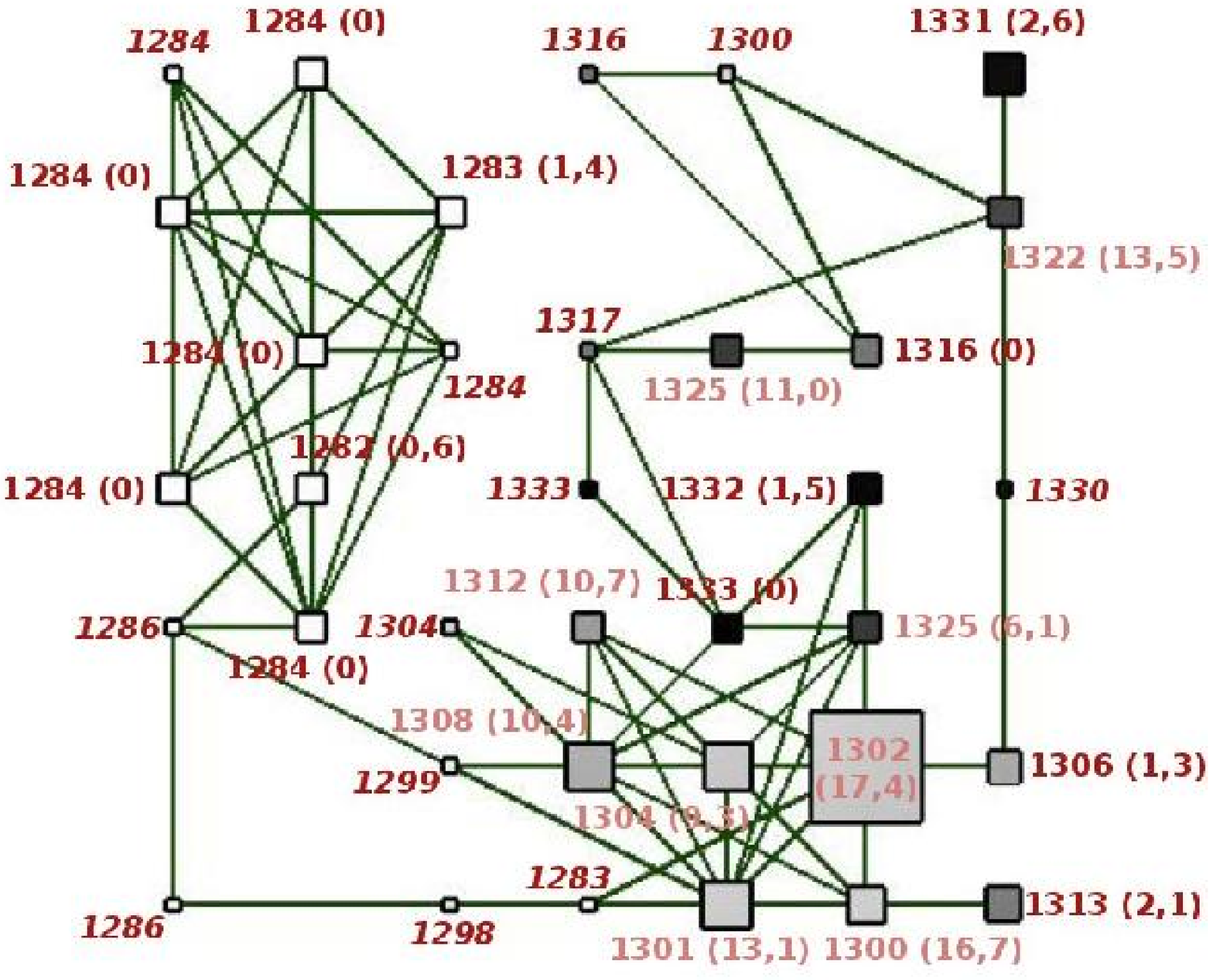} & \includegraphics[height= 5 cm]{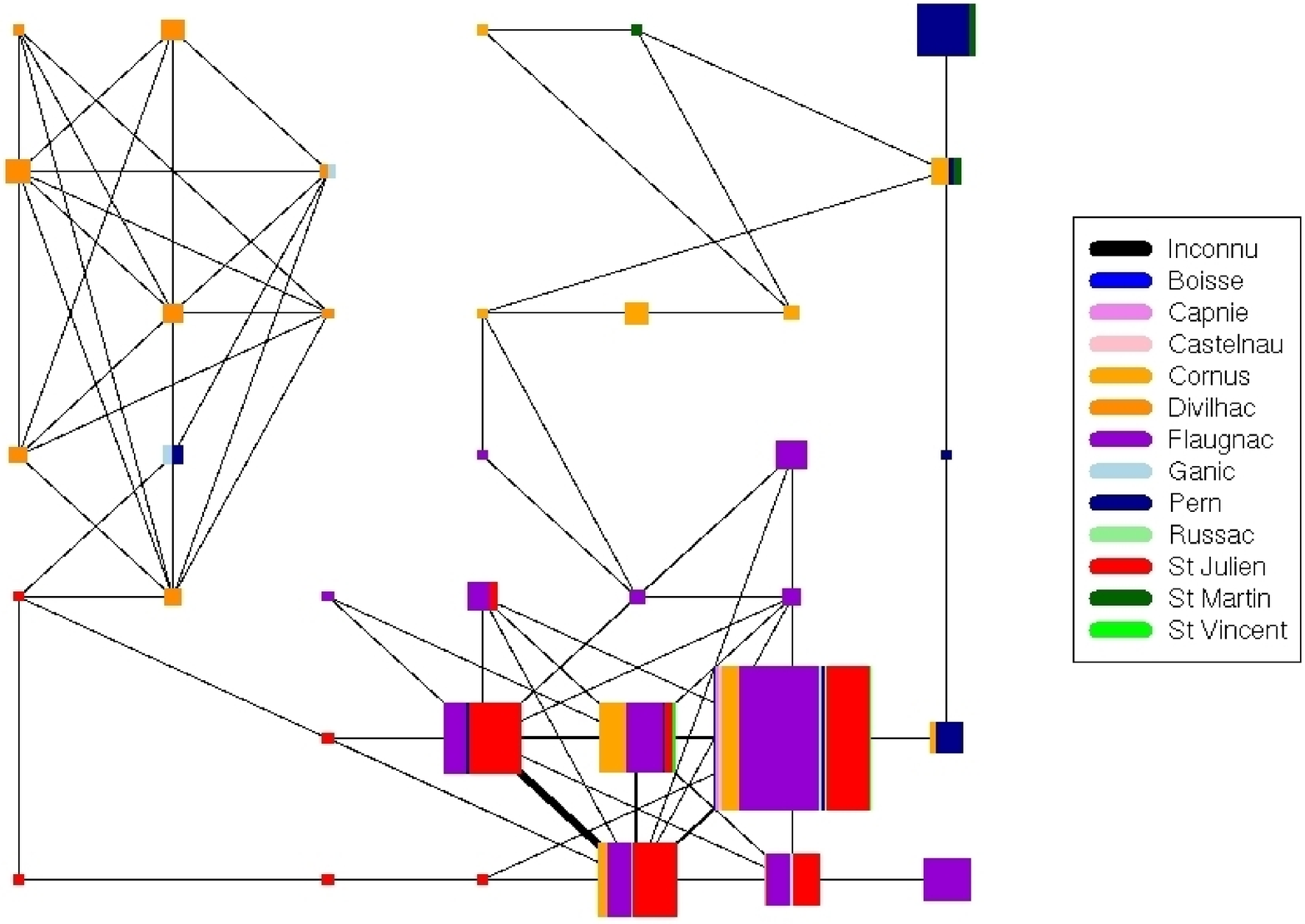}
		\end{tabular}
		\caption{Mean dates of each clusters with standard deviation in parenthesis (left) and geographical settings distribution of each cluster (right)\label{figure_date_et_lieu}}
	\end{center}
\end{figure}

This fact is clearly explained by Figure~\ref{figure_date_et_lieu} (left) where the significance of each cluster clearly appears: the top left part of the map is the oldest cluster whereas the top right part is the youngest, with a continuous connexion of the dates on the map. Figure~\ref{figure_date_et_lieu} (right) also provides some interesting informations about the social network: in particular, the top left part of the map has an homogeneous geographical setting which is the small village of Divilhac. This part of the map is only linked with the large clusters at the bottom right by a single peasant. This peasant doesn't live in this village but in the dominant village of the clusters to which he is linked at the bottom right of the map (St Julien\footnote{Readers interested by the location of the villages named in this paper will find a approximate map at \url{http://maps.google.com/maps/ms?ie=UTF8&hl=fr&msa=0&msid=100355826667676777753.000001134e74760eae6cd&z=10}}). Then, it seems that generational relationships and geographical ones are very important in this network. Moreover, it is suprising to see how even large clusters have a good homogeneity of their geographical settings (see the top right and bottom right clusters for example).

\section{Representing the whole graph from the kernel Self-Organizing Map\label{drawing_som}}

\begin{figure}[ht]
	\begin{center}
		\includegraphics[height=6 cm]{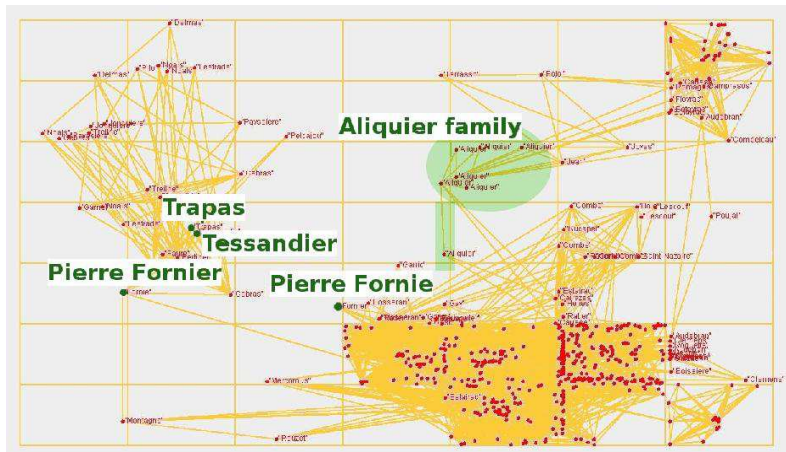}
		\caption{Representation of the whole medieval graph coming from kernel SOM\label{figure_ksom_entier}}
	\end{center}

\end{figure}

As the layout used for Figure~\ref{figure_ksom} has been built to be well organized, it provides an interesting starting point for a readable presentation of the whole graph. In \cite{truong_etal_SMESME2008,truong_etal_VSST2007}, Truong {\it et al.} developed an energy model, in the spirit of force directed algorithms, but under location constraints. This model intends to represent graphs that are already clustered. By applying this algorithm to the self-organizing map presented in Figure~\ref{figure_ksom}, we obtain the representation of Figure~\ref{figure_ksom_entier} where the names of some peasants in the smallest clusters have been added.

If it is obvious that the representation of the bottom right part of the map still has to be improved, some important facts that seem to be of interest for historians have been emphasized: first of all the peasant that links the top left part of the map to the bottom right one is ``Pierre Fornie''. This man is already known by historians to be a major character. This name also appear a bit appart from the main clusters and was identified by historians as the same people (and not a namesake) which means that some ambiguities still exist in the database (database correction is currently underway, partly due to this first analysis).

Moreover, the persons having a geographical setting different from the rest of the top left part of the map (Pern and Ganic instead of Divilhac) are named Trapas and Tessendier. They also belong to families known for their dominant positions. Then, some clusters roughly homogeneous on the geographical point of view are connected to similar clusters via important families that do not live in the same area and that can be seen as important links between villages. Finally, the top right part of the map is linked to the bottom right one by a single family named Aliquier family, this leads to identify this family as being very important for the social cohesion of the network.

All these remarks have helped historians to understand the organization of the social network. Moreover, dominant families have been identified through this first study: the next objective of this project is to understand how they structured the society and also how they have evolved through the hard break of the Hundred Years' war.

\vspace*{0.5 cm}

{\footnotesize
\baselineskip= 0.7pt

This work won't have existed without the ANR Graph-Comp's team. The authors thank Bertrand Jouve, project's coordinator, for this very interesting subject and for all discussions about it. We also want to thank Romain Boulet, Taofiq Dkaki and Pascale Kuntz for helpful discussions, Fabien Picarougne and Bleuenn Le Goffic who entirely created and managed the database and, of course, Florent Hautefeuille, historian at UMR TRACES (University of Toulouse Le Mirail), who provides us helpful comments and analysis.

}

\end{document}